\documentstyle[12pt,psfig]{article}
\begin{document}
\title{Production and absorption of $c \bar{c}$ pairs in nuclear
collisions at SPS energies\footnote{Supported by BMBF and GSI Darmstadt}}
\author{W. Cassing and E. L. Bratkovskaya \\
Institut f\"ur Theoretische Physik, Universit\"at Giessen \\
D-35392 Giessen, Germany}
\date{ }
\maketitle

\begin{abstract}
We study the production of $c \bar{c}$ pairs and dimuons from hard
collisions in nuclear reactions within the covariant transport approach
HSD, which describes successfully both hadronic and electromagnetic
observables from p~+~A and A~+~A collisions from SIS to SPS energies.
The production of $c \bar{c}$ and Drell-Yan pairs is treated
perturbatively employing experimental cross sections while the
interactions of $c\bar{c}$ pairs with hadrons are included by
conventional cascade-type two-body collisions. Adopting 6~mb for the $c
\bar{c}$-baryon cross sections the data on $J/\Psi$ suppression in
p~+~A reactions are reproduced in line with calculations based on the
Glauber model. We study different models for $c \bar{c}$ dissociation
on mesons in comparison with the experimental data of the HELIOS-3,
NA38 and NA50 collaborations. Adopting absorption cross sections with
mesons above the $D\bar{D}$ threshold in the order of 1.5 - 3~mb we
find that all data on $J/\Psi$ suppression from both proton-nucleus and
nucleus-nucleus collisions can be described without assuming the
formation of a quark-gluon plasma in these collisions.
\end{abstract}

\vspace{0.4cm}
\noindent
Nucl. Phys. A (1997), in print

\newpage
\section{Introduction}
The study of hot and dense nuclear matter via relativistic
nucleus-nucleus collisions is the major aim of high energy heavy-ion
physics. Nowadays, the search for a restoration of chiral symmetry at
high baryon density and temperature or for a phase transition to the
quark-gluon plasma (QGP) is of specific interest. In this context
Matsui and Satz \cite{matsui} have proposed that a suppression of the
$J/\Psi$ yield in ultra-relativistic heavy-ion collisions is a
plausible signature for the formation of the quark-gluon plasma because
the $J/\Psi$ should dissolve in the QGP due to color screening
\cite{matsui}.  This suggestion has stimulated a number of
heavy-ion experiments at CERN SPS to measure the $J/\Psi$ via its
dimuon decay. Indeed, these experiments have shown a significant
reduction of the $J/\Psi$ yield when going from proton-nucleus to
nucleus-nucleus collisions \cite{NA38}. Especially for Pb~+~Pb at
160~GeV/A an even more dramatic reduction of $J/\Psi$ has been reported
by the NA50 collaboration \cite{NA50,gonin}.

To interpret the experimental results, various models based on $J/\Psi$
absorption by hadrons have also been proposed. In Ref. \cite{gerschel},
Gerschel and H\"ufner have shown within the Glauber model that the
observed suppression of $J/\Psi$ in nuclear collisions is consistent
with the hadronic absorption scenario if one assumes a $J/\Psi$-nucleon
absorption cross section of about 6-7~mb.  Similar but more recent
analyses by the NA50 collaboration \cite{gonin} and Kharzeev \cite{KhaQM96}
have led to the same conclusion.  However, this model has failed to
explain the ``anomalous" suppression reported in central Pb~+~Pb
collisions, thus leading to the suggestion of a possible formation of a
quark-gluon plasma in these collisions \cite{gonin,KhaQM96,blaizot,wong}.
On the other hand, Gavin et al. \cite{gavin,gavin1}, based also
on the hadronic absorption model, have found that although $J/\Psi$
absorption by nucleons is sufficient to account for the measured total
$J/\Psi$ cross sections in both proton-nucleus and nucleus-nucleus
collisions, it cannot explain the transverse energy dependence of
$J/\Psi$ suppression in nucleus-nucleus collisions.  To account for the
nucleus-nucleus data they have introduced additionally the absorption
on mesons (`comovers') with a cross section of about 3~mb. A similar
model has also been proposed by Capella et al. \cite{Capella} to describe
the $J/\Psi$ and $\Psi^\prime$ suppression in nucleus-nucleus
collisions. On the other hand, Kharzeev et al.  \cite{KLNS96} claim the
`comover' absorption model to be inconsistent when considering all data
on $J/\Psi$ production simultaneously.

In all these studies the dynamics of the collisions is based on the
Glauber model, so a detailed space and time evolution of the colliding
system is not included. In particular, the transverse expansion of the
system and the finite hadron formation time is ignored in the Glauber
models.  Especially for nucleus-nucleus collisions involving heavier
beams, such as the Pb~+~Pb collisions at 160~GeV/nucleon, the dynamics is
more subtle than in proton and S induced reactions. Thus dynamical models
are needed to complement our information on the reaction dynamics.
In this respect Loh et al. \cite{loh} have investigated the $J/\Psi$
dissociation in a color electric flux tube in a semiclassical model based
on the Friedberg-Lee color dielectric Lagrangian. They find that the
$c\bar{c}$ dissociation time is in the order of 1 fm/c. A first transport
theoretical analysis of $J/\Psi$ production and absorption has been
performed in Ref. \cite{Ca97} where the $c \bar{c}$ production was
based on the LUND string formation and fragmentation model \cite{LUND}.
Indeed, substantial differences to the Glauber approaches have been
found due to a finite formation time of the $c \bar{c}$ pair and its
subsequent interactions with baryons and mesons.  However, due to the
low statistics achieved in the numerical calculations a definite
conclusion about the charmonium suppression could not be obtained in
the latter study since alternative production and absorption schemes
also could lead to different absorption rates.

In this work we continue the studies in Ref.~\cite{Ca97} on $c \bar{c}$
production and suppression at SPS energies within the covariant
transport approach HSD\footnote{Hadron String Dynamics} \cite{cassing}
using different production and absorption models.  The nonequilibrium
transport calculations have been shown to describe satisfactorily both
the measured hadronic observables (rapidity distributions, transverse
momentum spectra etc.  \cite{cassing}), which are sensitive to the
final stage collision dynamics, and dilepton spectra
\cite{cassing1,cassing2,brat96,brat97}, which reflect also the initial
hot dense stage of the collisions. It thus gives a more realistic
description of the heavy-ion reaction dynamics than that used in Refs.
\cite{gonin,gerschel,KhaQM96,blaizot,gavin,gavin1}. Within this
approach we can check if the $c \bar{c}$ pair might be destroyed by
nucleons before the mesons are produced as argued in Ref.
\cite{blaizot} (cf. also Refs.~\cite{KhaQM96,wong}). Furthermore, we
can test if a finite lifetime of the $c \bar{c}$ pre-resonance system
as suggested by Kharzeev et al. \cite{KhaQM96,KhSatz96} comes in
conflict with the data since according to Ref.  \cite{kharzeev} the
$J/\Psi$ - meson cross section might be negligibly small in hadronic
matter due to the small size of the $J/\Psi$ and its large mass gap from 
open charms. However, it is expected that the $c\bar{c}$ pair is
first produced in a color-octet state together with a gluon
(`pre-resonance state') and that this more extended configuration has a
larger interaction cross section with baryons and mesons before the
$J/\Psi$ singlet state finally emerges.  We will not address the
question of whether the magnitude of the $c \bar{c}$-hadron cross
sections used is correct or can be justified by nonperturbative QCD and
concentrate on the question if specific reactions models can be ruled
out in comparison to the data available so far.

The outline of the paper is as follows: In Section 2 we briefly describe
the covariant transport approach employed and present details of the
baryon and meson dynamics for S~+~U at 200~GeV/A and Pb~+~Pb at 160~GeV/A.
Section 3 contains the description of charmonium and Drell-Yan production
by nucleon-nucleon collisions as well as two models for the charmonium
reabsorption on hadrons. In Section 4 we present a detailed comparison of
our calculations with the experimental data available so far while Section 5
concludes with a summary and discussion of open problems.

\section{The covariant transport approach}
In this work we perform our analysis along the line of the HSD approach
\cite{cassing} which is based on a coupled set of covariant transport
equations for the phase-space distributions $f_{h} (x,p)$ of hadron $h$
\cite{cassing,Weber1}, i.e.
\begin{eqnarray}  \label{g24}
\lefteqn{\left\{ \left( \Pi_{\mu}-\Pi_{\nu}\partial_{\mu}^p U_{h}^{\nu}
-M_{h}^*\partial^p_{\mu} U_{h}^{S} \right)\partial_x^{\mu}
+ \left( \Pi_{\nu} \partial^x_{\mu} U^{\nu}_{h}+
M^*_{h} \partial^x_{\mu}U^{S}_{h}\right) \partial^{\mu}_p
\right\} f_{h}(x,p) } \nonumber \\
&& = \sum_{h_2 h_3 h_4\ldots} \int d2 d3 d4 \ldots
 [G^{\dagger}G]_{12\to 34\ldots}
\delta^4(\Pi +\Pi_2-\Pi_3-\Pi_4 \ldots )  \nonumber\\
&& \times \left\{ f_{h_3}(x,p_3)f_{h_4}(x,p_4)\bar{f}_{h}(x,p)
\bar{f}_{h_2}(x,p_2)\right.  \nonumber\\
&& -\left. f_{h}(x,p)f_{h_2}(x,p_2)\bar{f}_{h_3}(x,p_3)
\bar{f}_{h_4}(x,p_4) \right\} \ldots\ \ .
\end{eqnarray}
In Eq.~(\ref{g24}) $U_{h}^{S}(x,p)$ and $U_{h}^{\mu}(x,p)$ denote the
real part of the scalar and vector hadron selfenergies, respectively,
while $[G^+G]_{12\to 34\ldots} \delta^4_{\Gamma} (\Pi
+\Pi_2-\Pi_3-\Pi_4\ldots )$ is the `transition rate' for the process
$1+2\to 3+4+\ldots$ which is taken to be on-shell in the
semiclassical limit adopted. The hadron quasi-particle properties in
(\ref{g24}) are defined via the mass-shell constraint \cite{Weber1},
\begin{equation}   \label{g25}
\delta (\Pi_{\mu}\Pi^{\mu}-M_{h}^{*2} ) \ \ ,
\end{equation}
with effective masses and momenta given by
\begin{eqnarray} \label{g26}
M_{h}^* (x,p)&=&M_h + U_h^{{S}}(x,p) \nonumber \\
\Pi^{\mu} (x,p)&=&p^{\mu}-U^{\mu}_h (x,p)\ \ ,
\end{eqnarray}
while the phase-space factors
\begin{equation}
\bar{f}_{h} (x,p)=1 \pm f_{{h}} (x,p)
\end{equation}
are responsible for fermion Pauli-blocking or Bose enhancement,
respectively, depending on the type of hadron in the final/initial
channel. The dots in Eq.~(\ref{g24}) stand for further contributions to
the collision term with more than two hadrons in the final/initial
channels. The transport approach (\ref{g24}) is fully specified by
$U_{h}^{S}(x,p)$ and $U_{h}^{\mu}(x,p)$ $(\mu =0,1,2,3)$, which
determine the mean-field propagation of the hadrons, and by the
transition rates $G^\dagger G\,\delta^4 (\ldots )$ in the collision
term, that describes the scattering and hadron production/absorption
rates.

The scalar and vector mean fields $U_{h}^{S}$ and $U^\mu_{h}$ for
baryons are taken from Ref.~\cite{cassing} and don't have to be
specified here again, since variations in the baryon selfenergies
within the constraints provided by experimental data were found to have
no sizeable effect on the issue of $c \bar{c}$ production and absorption.
In the present approach we propagate explicitly -- apart from the
baryons (cf.~\cite{cassing}) -- pions, kaons, $\eta$'s, $\eta^\prime$'s,
the $1^-$ vector mesons $\rho, \omega, \phi$ and $K^*$'s as well as the
axial vector meson $a_1$. The production of mesons and baryon-antibaryon
pairs is treated within the LUND string model \cite{LUND} employing a
formation time $\tau_F$ = 0.7 fm/c which controls the baryon and meson
rapidity distributions $dN/dy$ in comparison to experimental data. As
meson-meson channels we include the reactions $\pi \pi \rightarrow \rho,
\pi \pi \rightarrow K \bar{K}, \pi \rho \rightarrow \phi, \pi \rho
\rightarrow a_1$ as well as the time reversed reactions using Breit-Wigner
cross sections with parameters from the literature \cite{PDB} and
exploiting detailed balance. As noted before, this transport approach
was found to describe reasonably well hadronic as well as dilepton
data from SIS to SPS energies \cite{cassing,brat96,brat97}.

\begin{figure}[h1]
\phantom{a}
\vspace*{-2cm}
{\psfig{figure=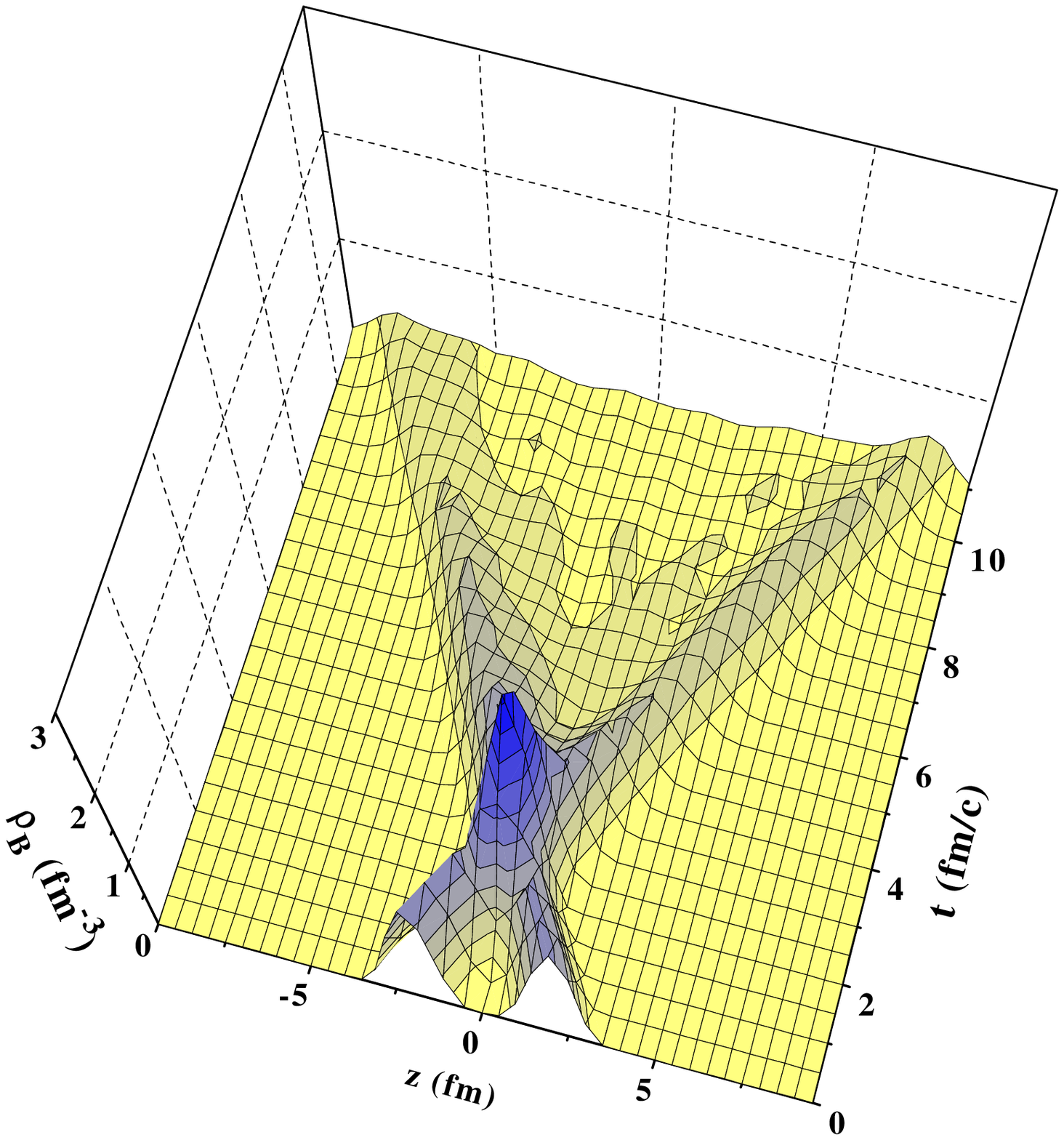,width=15cm,height=22cm}}
\vspace*{-3cm}
\caption{The baryon density $\rho_B(x=0,y=0,z;t)$ for a Pb~+~Pb
collision at 160~GeV/A and impact parameter $b = 1$~fm.}
\label{Fig1}
\end{figure}

Before discussing the question of charmonium production we show in
Fig.~\ref{Fig1} the time evolution of the baryon density $\rho_B(x,y,z;t)$
as a function of $z$ and time $t$ for $x=y=0$ in a central collision of
Pb~+~Pb at 160~GeV/A in the nucleon-nucleon center-of-mass frame. The
two Pb-ions start to overlap at $t \approx$ 1 fm/c, get compressed up
to a maximum density of about 2.5 fm$^{-3}$ at $t \approx$ 2.5 fm/c and
expand in longitudinal ($z$) direction later on indicating a sizeable
amount of transparency. We note that the space-time evolution in
Fig.~\ref{Fig1} is controlled by the experimental rapidity
distributions $dN/dy$ for protons and negatively charged particles from 
NA49 \cite{NA49}; the streaming of hadrons or more precisely their
distribution in velocity $\beta$, i.e. $dN/d\beta$ can also be directly
extracted from the experimental data using $dN/d\beta = \frac{1}{2}
dN/dy(\beta) \ (1/(1 + \beta) + 1/(1 - \beta))$.

\begin{figure}[h1]
\vspace*{-2cm}
{\psfig{figure=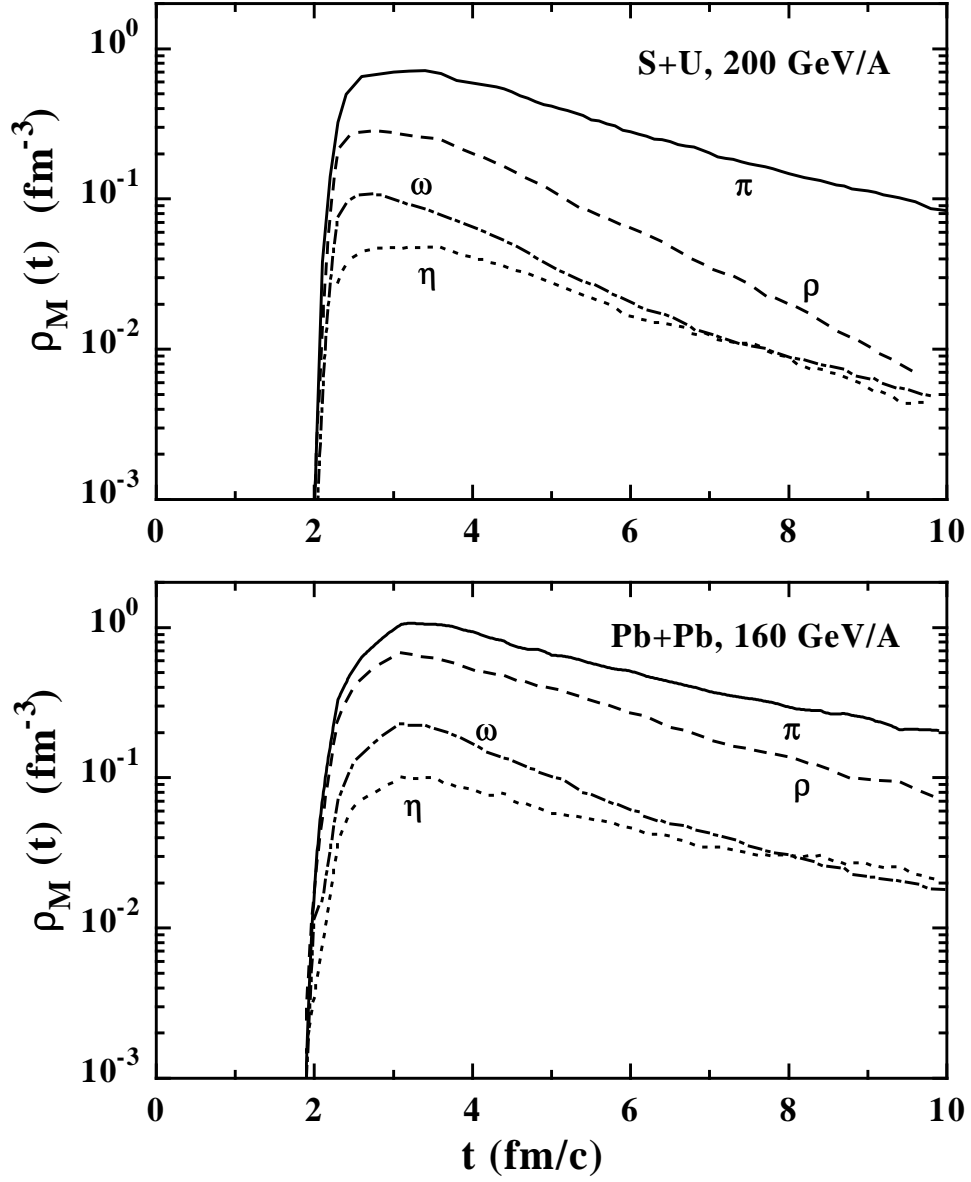,width=15cm,height=22cm}}
\vspace*{-2cm}
\caption{The density of $\pi, \rho, \omega$ and $\eta$ mesons in a
central cylinder of radius $R = 3$~fm and volume $V \approx 15\ {\rm fm}^3$
for a S~+~U collision at 200~GeV/A and a Pb~+~Pb collision at 160~GeV/A
for $b = 1$~fm.}
\label{Fig2}
\end{figure}

The produced mesons in this reaction (in a central cylinder of radius
$R=3$~fm and volume  $V \approx 15 \ {\rm fm}^3$) appear at about
$t \approx$ 2 fm/c as can be extracted from Fig.~\ref{Fig2} (lower part)
where the densities of pions, $\rho, \omega$ and $\eta$ mesons are
displayed separately as a function of time. The maximum in the meson
density ($\approx 1 \ {\rm fm}^{-3}$ for pions) at $t \approx$ 3.2 fm/c
appears with a delay of $\tau_F$ = 0.7 fm/c with respect to the maximum
baryon density (cf. Fig.~\ref{Fig1}). For comparison we also show in
Fig.~\ref{Fig2} the meson densities for a central S~+~U collision at
200~GeV/A within the same volume. When comparing to the central Pb~+~Pb
collision at 160~GeV/A we observe a lower meson density for the S + U
case in the central overlap region; especially the $\rho$-meson
density is lower by about a factor of 2. Nevertheless, high baryon and
meson densities are encountered in these reactions for t $\geq$ 2 fm/c
where a $c \bar{c}$ pair -- that can be produced from 1 - 3.5 fm/c --
has to pass through.

\begin{figure}[h1]
\vspace*{-2cm}
{\psfig{figure=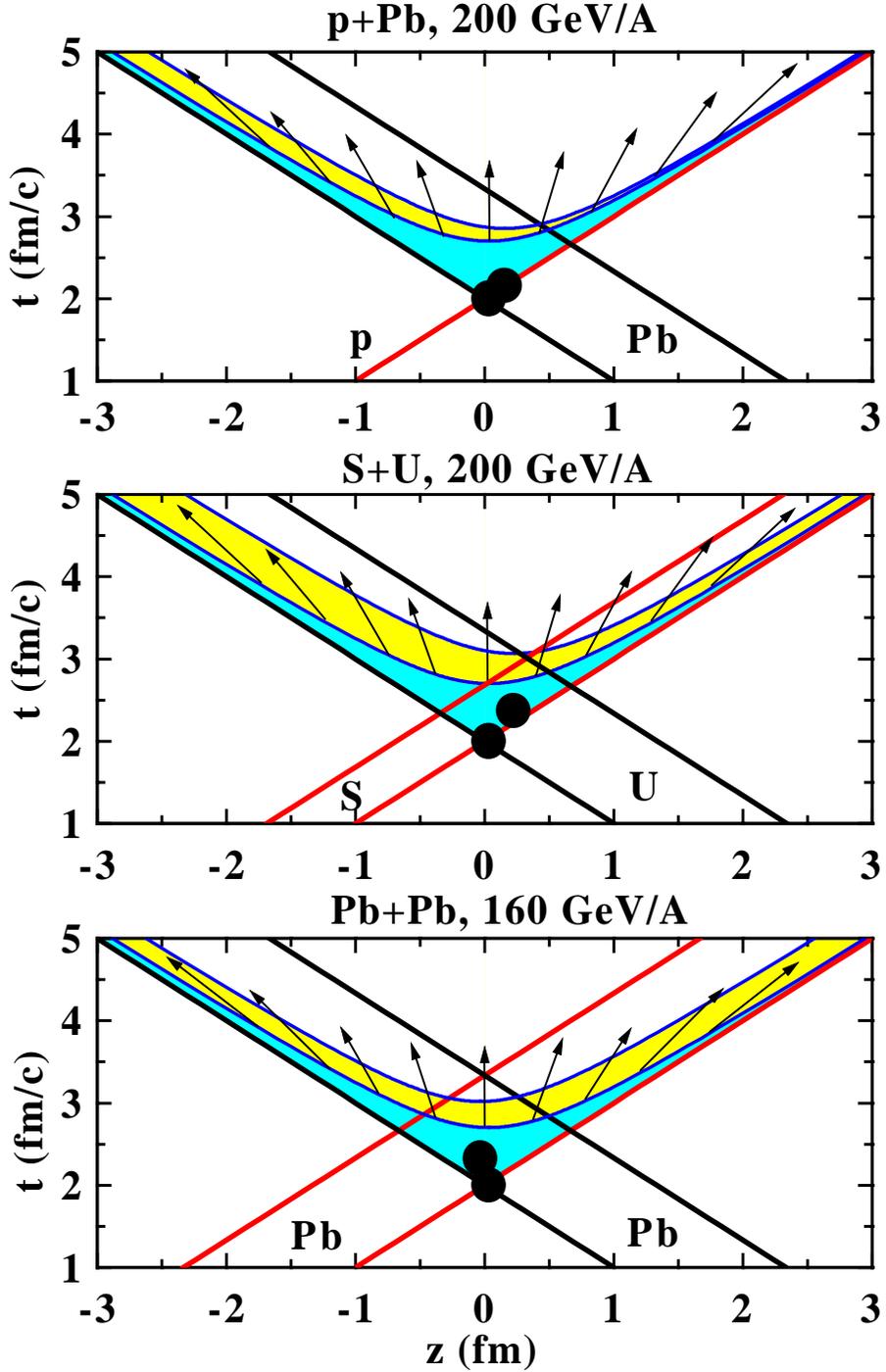,width=15cm,height=22cm}}
\vspace*{-2cm}
\caption{Schematic representation of p~+~Pb collision at 200~GeV (upper
part), S~+~U collision at 200~GeV/A (middle part) and a Pb~+~Pb
collision at 160~GeV/A (lower part) in space-time.  The full dots
represent early hard collision events (for Drell-Yan and $c\bar{c}$
pairs) while mesons ($\pi, \eta, \rho, \omega$, etc. -- arrows) only
appear after a respective formation time $\tau_F \approx$ 0.7~fm/c.
The overlap area (inner rectangle) specifies the space-time region of
hard production events.}
\label{Fig3}
\end{figure}

This situation is summarized schematically in Fig.~\ref{Fig3} for a p +
Pb (upper part) and S~+~U collision at 200~GeV (middle part) as well as
for a central Pb~+~Pb collision at 160~GeV/A (lower part) for freely
streaming baryons (thick lines). In all cases the initial string
formation space-time points are indicated by the full dots; the mesons
(indicated by arrows) hadronize after a time delay $\tau_F$ = 0.7 fm/c
as shown by the first hyperbola.
A $c \bar{c}$ pair produced in the initial
hard nucleon-nucleon collision cannot be absorped by mesons in the dark
shaded areas in space and time; however, in case of Pb~+~Pb, where
$c\bar{c}$ pairs should be produced within the inner rectangles, a
sizeable fraction will also be produced in a dense mesonic environment.
This fraction of pairs produced at finite meson density for S~+~U is
much reduced as can be seen from Fig.~\ref{Fig3} (middle part).

The upper hyperbolas in Fig. 3 represent the boundaries for the appearance
of mesons from the second interaction points (full dots) which appear
somewhat later in time; they stand for a representative further nucleon -
nucleon collision during the reaction.

\section{Charmonium and Drell-Yan production}

\begin{figure}[h1]
\vspace*{-2cm}
{\psfig{figure=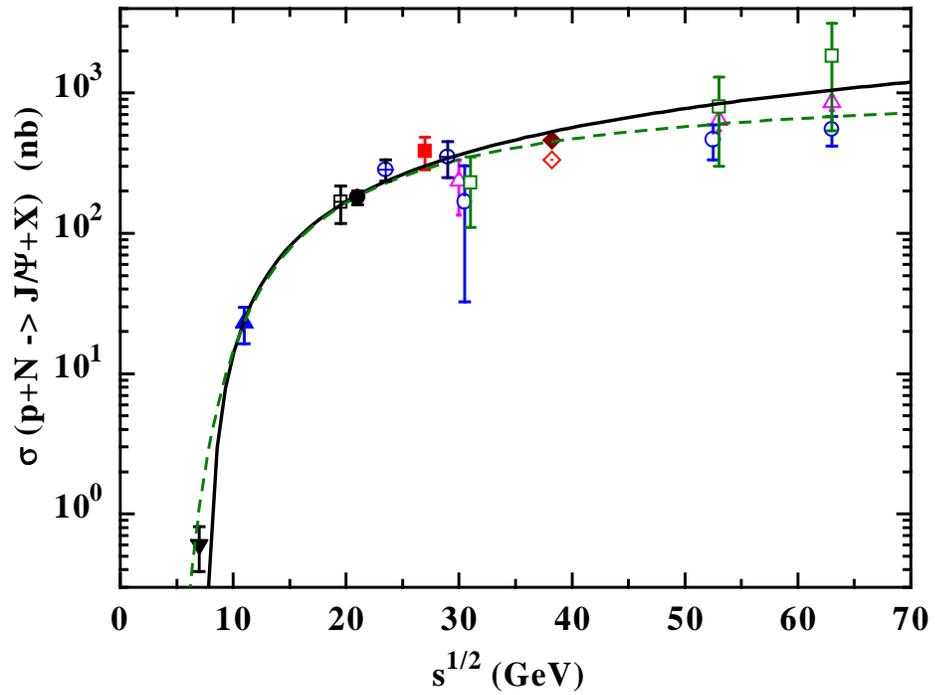,width=15cm,height=22cm}}
\vspace*{-2cm}
\caption{The parametrization (\protect\ref{selem}) for the elementary
$J/\Psi$ cross section in pp collisions (solid line) in comparison to
the experimental data from Ref.~\protect\cite{CarQM96}. The dashed line
represents the parametrization (\protect\ref{scarlos}).}
\label{Fig4}
\end{figure}

Since the probability of producing a $c \bar{c}$ or Drell-Yan pair is
very small, a perturbative approach is used for technical reasons.
Whenever  two nucleons collide a $c\bar{c}$ pair is produced with a
probability factor $W$, which is given by the ratio of the $J/\Psi$ (or
$\Psi^\prime$) to $NN$ cross section at a center-of-mass energy
$\sqrt{s}$ of the baryon-baryon collision,
\begin{equation}
W= \frac{\sigma_{BB \rightarrow J/\Psi + X} (\sqrt{s})}
{\sigma_{BB \rightarrow BB+X} (\sqrt{s})}.
\label{wratio}
\end{equation}
The parametrization used for the $J/\Psi$ cross section is
\begin{equation}
\sigma_{BB \rightarrow J/\Psi + X} (\sqrt{s}) = 160 \
\left(1 - \frac{a}{\sqrt{s}}\right)^2 \ \frac{\sqrt{s}}{a} \ [nb]
\label{selem}
\end{equation}
with $a$ = 7.47~GeV which is shown in Fig.~\ref{Fig4} (solid line) in
comparison to the experimental data displayed in Ref. \cite{CarQM96}.
For the systems to be studied, in the energy regime $\sqrt{s} \leq$
30~GeV, the estimated uncertainty of our parametrization is about
20~\%. We note that the parametrization used conventionally
\cite{CarQM96} is
\begin{eqnarray}
\sigma_{pp \rightarrow J/\Psi + X} (\sqrt{s}) = d \ \left( 1 -
{c\over \sqrt{s}}\right)^{12}
\label{scarlos}\end{eqnarray}
with $c=3.097$~GeV, $d=2 \cdot 37 / B_{\mu\mu}$~nb while
$B_{\mu\mu}=0.0597$ is the branching ratio of the $J/\Psi$  to dimuons.
The expression (\ref{scarlos}) (dashed line in Fig.~\ref{Fig4}) is in a
good agreement with our formula (\ref{selem}) for 10~GeV $\le \sqrt{s}
\le$ 30~GeV which is the energy regime of interest. The rapidity
distribution of the $c \bar{c}$-pair is approximated by a Gaussian in
the nucleon-nucleon center-of-mass of width $\sigma \approx$ 0.6 as in
Ref.~\cite{Carlos} while the transverse momentum distribution is fitted
to experimental data (see below). For $\Psi^\prime$ production we
employ the same model, however, scale the experimental cross section by
a factor of 0.122 relative to the $J/\Psi$ production cross section.

In extension to Ref.~\cite{Ca97} the Drell-Yan process is taken into
account explicitly. The generation of Drell-Yan events was performed
with the PYTHIA event generator \cite{PYTHIA} version 5.7 using GRV LO
\cite{GRVLO} or MRS A \cite{MRSA} structure functions from the PDFLIB
package \cite{PDFLIB} with $k_T = 1.0$~GeV.  In the kinematical
domain $1.5 \le M \le 5.5$~GeV, $3\le y_{lab}\le 4$ and $-0.5 \le
\cos\theta_{CS} \le 0.5$, where $\theta_{CS}$ is the polar angle in the
Collins-Soper reference frame, this yields a dimuon cross section of 270
pb in pp collisions at 200~GeV and 261~pb in pn collisions,
respectively, as in Refs.~\cite{NA38,Carlos}.

According to our dynamical prescription the Drell-Yan pairs can be
created in each hard $pp, pn, np$ or $nn$ collision ($\sqrt{s} \geq$
10~GeV).  Since PYTHIA calculates the Drell-Yan process in leading
order only (using GRV LO structure functions) we have multiplied the
Drell-Yan yield for $NN$ collisions by a K-factor of 2.0 (cf. Refs.
\cite{NA38,Carlos}).  In case of MRS A structure functions, which
include NLO corrections, a K-factor of about 1.6 had to be introduced
(cf. \cite{CarQM96}).  The energy distribution of the hard $NN$
collisions $dN/d\sqrt{s}$ for nucleus-nucleus collisions in our
transport approach shows a pronounced peak around $\sqrt{s_0} = \sqrt{2 m_N
(T_{kin} + 2 m_N)}$ where $T_{kin}$ is the kinetic energy per nucleon in
the laboratory frame.  Thus the main contribution to the dimuon yield --
summed over all $NN$ events -- comes from $NN$ collisions with
$\sqrt{s}\simeq \sqrt{s_0}$.  However, there are also Drell-Yan pairs from 
$NN$ collisions with $\sqrt{s}$ larger or smaller than
$\sqrt{s_0}$.

We have compared our results within the production scheme given above
with that used in Refs. \cite{NA38,NA50,Carlos}, where the Drell-Yan
yield from p~+~A and A~+~A collisions is calculated as the isotopical
combination of the yield from $pp$ and $pn$ at fixed $\sqrt{s_0}$
scaled by $A_P \times A_T$. We found that the variation from the scheme
used in Refs. \cite{NA38,NA50,Carlos} is less then 10\%.  We,
furthermore, note that the difference in the dimuon spectrum using
different structure functions (GRV LO or MRS A) is less then 5\%.
In the present analysis we have discarded dimuon production from open
charm channels because a recent analysis by Braun-Munzinger et al.
\cite{BM} on the basis of the same PYTHIA event generator has shown
that the open charm contributions at low and high invariant masses
are of minor importance.

Since the production scheme is the same for
$c\bar{c}$-pairs\footnote{The formation time $\tau_F$ for the
$c\bar{c}$-pair is assumed to be zero in our present approach in contrast
to Ref. \cite{Ca97}.} their
total cross section (without reabsorption) also scales with $A_P \times
A_T$; the ratio of the $J/\Psi$ to the Drell-Yan cross section thus
provides a direct measure for the $J/\Psi$ suppression.

\begin{figure}[h1]
\vspace*{-2cm}
{\psfig{figure=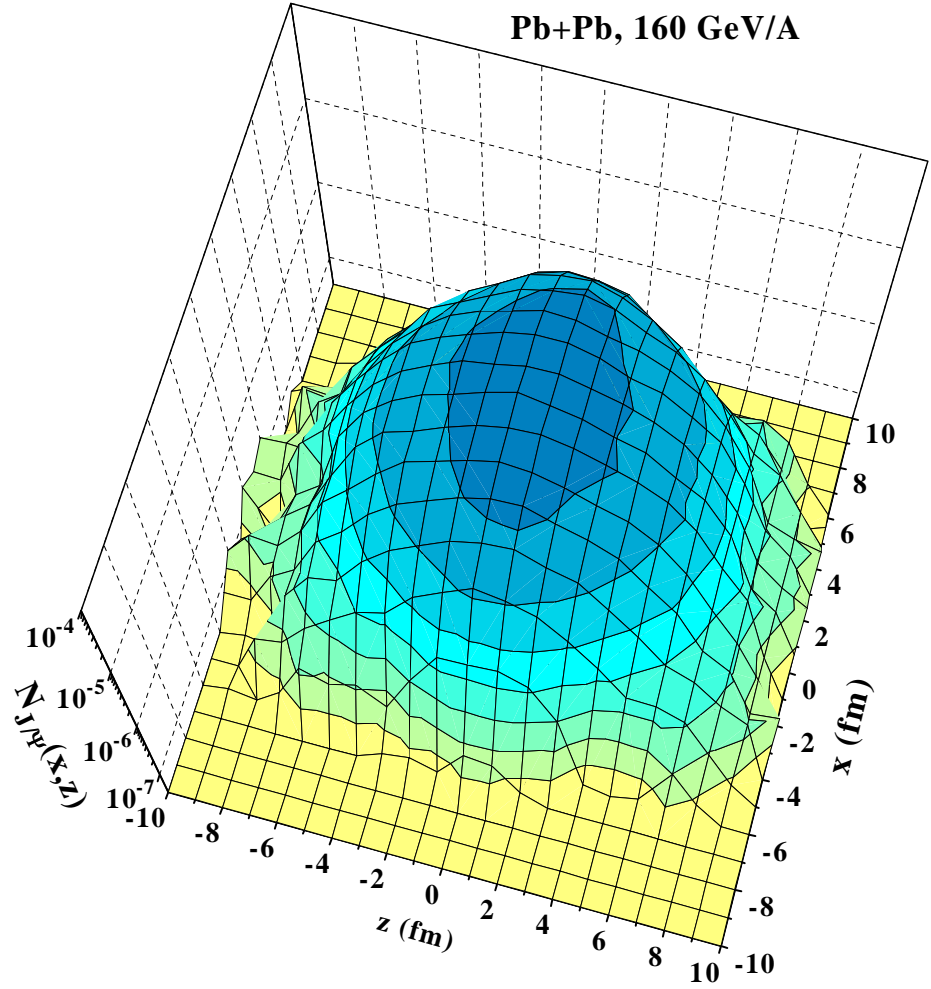,width=15cm,height=22cm}}
\vspace*{-2cm}
\caption{The production probability for $c\bar{c}$-pairs in a central
Pb~+~Pb collision at 160~GeV/A in the $(x,z)$-plane integrated over
$y$. The $z$-axis is scaled by the Lorentz factor $\gamma_{cm}$ = 9.3
to compensate for the Lorentz contraction in beam direction.}
\label{Fig5}
\end{figure}

In order to obtain some information about the primary distribution of
the produced `pre-resonance' states in coordinate space we show in
Fig.~\ref{Fig5} the $c\bar{c}$ distribution in the $(x,z)$-plane
integrated over y for a central collision of Pb~+~Pb at 160~GeV/A. The
collision of the two nuclei proceeds along the $z$-direction and the
actual $z$-axis has been stretched by the Lorentz factor $\gamma_{cm}
\approx 9.3$ to compensate for the Lorentz contraction in beam
direction.  It is clearly seen from Fig.~\ref{Fig5} that the production
of the charmonium state by hard nucleon-nucleon collisions is enhanced
in the center and drops rapidly in the surface region of the
overlapping nuclei.

We follow the motion of the $c \bar{c}$ pair in hadronic matter
throughout the collision dynamics by propagating it as a free particle.
In our simulations the $c\bar{c}$ pair, furthermore, may be destroyed
in collisions with hadrons using the minimum distance concept as
described in Sec. 2.3 of Ref.~\cite{Wolf90}.  For the actual cross
sections employed we study two models (denoted by I and II) which both
assume that the $c\bar{c}$ pair initially is produced in a color-octet
state and immediately picks up a soft gluon to form a color neutral
$c\bar{c}-g$ Fock state \cite{KhaQM96} (color dipole).  This extended
configuration in space is assumed to have a 6~mb dissociation cross
section in collisions with baryons ($c \bar{c}+B\to\Lambda_c+\bar D$)
as in Refs.  \cite{gonin,gerschel,KhaQM96} during the lifetime $\tau$
of the $c\bar{c}-g$ state which is a parameter. In the model I we
assume $\tau = $ 10~fm/c which is large compared to the nucleus-nucleus
reaction time such that the final resonance states $J/\Psi$ and
$\Psi^\prime$ are formed in the vacuum without further interactions
with hadrons. In the model II we adopt $\tau =$ 0.3 fm/c as suggested
by Kharzeev \cite{KhaQM96} which implies also to specify the
dissociation cross sections of the formed resonances $J/\Psi$ and
$\Psi^\prime$ on baryons. For simplicity we use 3~mb following Ref.
\cite{KhSatz96}.  The cross section for $c\bar{c}-g, J/\Psi$, or
$\Psi^\prime$ dissociation on mesons ($c\bar{c}+m\to D\bar{D}$) is
treated as a free parameter ranging from 0 to 3~mb.

\begin{figure}[h1]
\vspace*{-2cm}
{\psfig{figure=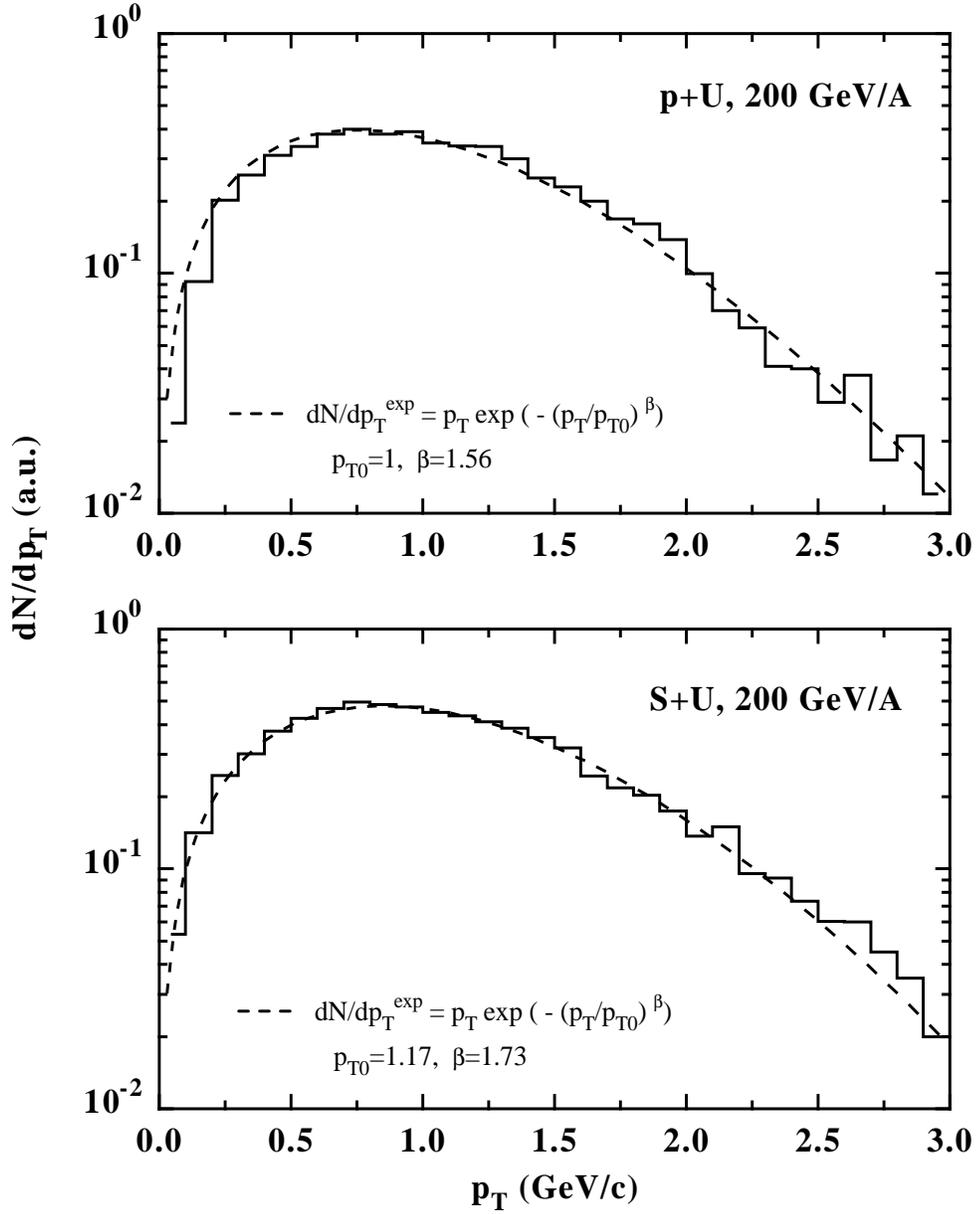,width=15cm,height=22cm}}
\vspace*{-2cm}
\caption{The transverse momentum distribution of $J/\Psi$ mesons in p~+~U 
(upper part) and S~+~U collisions (lower part) at 200~GeV/A from 
the HSD calculations (histograms). The solid lines are fits to the
experimental data from Ref.~\protect\cite{NA38}.}
\label{Fig6}
\end{figure}

In order to test the transverse momentum dependence of the $J/\Psi$
production we have performed calculations for p~+~U and S~+~U at 200
GeV/A (using model I with $\sigma_{abs}^{baryons}$ = 6~mb and
$\sigma_{abs}^{mesons}$ = 1~mb) which are scaled in magnitude to the
respective data for the same systems from Ref.~\cite{NA38}. Our
calculated results are shown in Fig.~\ref{Fig6} for both systems in
terms of the histograms while the solid lines represent fits to the
experimental transverse momentum distributions. Since this $p_T$
dependence is described fairly well we expect the $c\bar{c}$ event
distributions to be quite realistic.

\section{Analysis of experimental data}

\subsection{HELIOS-3 data}
Since the $J/\Psi$ and $\Psi^\prime$ are measured in nuclear collisions
through their decay into dimuons, we calculate explicitly the dimuon
invariant mass spectra from these collisions.  This includes not only
the decay of the $J/\Psi, \Psi^\prime$ but also the decay of other
mesons ($\eta$, $\rho$, $\omega$, and $\phi$) as well as the Dalitz
decay of the $\eta$, $\omega$, $\eta^\prime$ etc.  Details on
calculating the dimuon spectra from heavy-ion collisions up to
invariant masses of about 1.3~GeV can be found in Refs.
\cite{cassing1,cassing2,brat97}.  Since the Drell-Yan contribution is
important for dileptons with invariant masses above 1.5~GeV
\cite{Carlos} we have included these channels by computing for each
`hard' nucleon-nucleon collision ($\sqrt{s} \geq$ 10~GeV) their
contribution via PYTHIA 5.7 \cite{PYTHIA} as described above.

\begin{figure}[h1]
\vspace*{-2.5cm}
{\psfig{figure=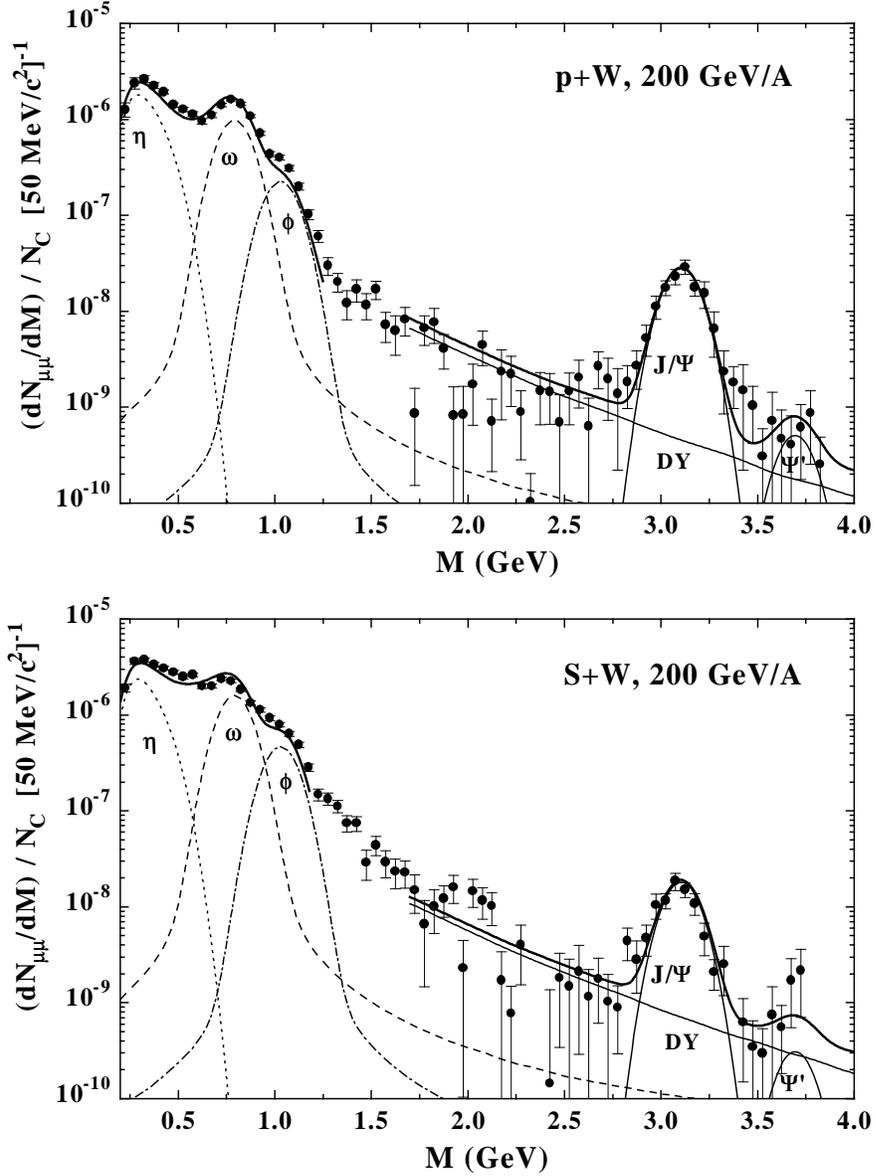,width=15cm,height=22cm}}
\vspace*{-3cm}
\caption{Dimuon invariant mass spectra from p~+~W and S~+~W collisions
at 200~GeV/nucleon in comparison to the data of the HELIOS-3
collaboration \protect\cite{helios}. In the low mass region the
individual contributions from the $\eta$ Dalitz decay, $\omega$ and
$\Phi$ decay are shown by the thin lines. The Drell-Yan contribution
(denoted by DY) is calculated only for invariant masses $M
\geq$~1.5~GeV. The explicit contributions from $J/\Psi$ and
$\Psi^\prime$ decays are indicated by thin lines for $M \geq$ 2.7~GeV.
The thick solid lines represent the sum of all dimuon channels (except
for open charm contributions). }
\label{Fig7}
\end{figure}

We have carried out calculations for p~+~W and central S~+~W collisions
at 200~GeV/A as in Ref. \cite{Ca97} in order to check on an absolute
scale if the production and absorption schemes employed in the HSD
approach are included properly. In Fig.~\ref{Fig7} we show the dilepton
invariant mass spectra for p + W and S + W at 200~GeV/A normalized to
the number of charged particles in the pseudorapidity bin 3.7 $\leq
\eta \leq$ 5.2 and compare them with the experimental data from Ref.
\cite{helios} including their acceptance and mass resolution. Again we
have employed model I with $\sigma_{abs}^{baryons}$ = 6~mb and
$\sigma_{abs}^{mesons}$ = 1~mb (see below).  It is seen from
Fig.~\ref{Fig7} that in the p~+~W and S~+~W cases the theoretical
results agree well with the data on an absolute scale which implies
that apart from the low mass dimuon spectrum - which has been analyzed
in Ref.~\cite{cassing2} - also the $J/\Psi$ and $\Psi^\prime$ region is
described reasonably well. At low dimuon masses the explicit
contributions from the mesons $\eta, \omega, \phi$ are displayed in
terms of the thin lines while for invariant masses above 1.5~GeV the
Drell-Yan (DY), $J/\Psi$ and $\Psi^\prime$ contributions are shown
explicitly. The full solid curve represents the sum of all
contributions (without open charm channels). It is intersected from $M
\approx $ 1.2~GeV to 1.5 GeV because the continuation of the Drell-Yan
contribution to lower energies is not clear as well as the tail of the
$\rho$-meson contribution at higher $M$.  Futhermore, open charm
channels (from $D$ and $\bar{D}$ mesons) have not been included yet
(cf. Ref. \cite{BM}). We note that for S~+~W in the invariant mass
range from 1.6~GeV~$\leq M \leq$~2.5~GeV the dimuon yield is almost
twice that for p + W within the present normalization and that our
calculations reproduce also the intermediate mass range on the basis of
PYTHIA 5.7 reasonably well.  Thus our calculations can be regarded as
fully microscopic studies on dilepton production for invariant masses
0.2~GeV~$\leq M \leq$~5~GeV including both the relevant `soft' and
`hard' processes.

\subsection{Proton-nucleus reactions}
Since the absorption of $c\bar{c}$-pairs on secondary mesons in
proton-nucleus collisions is practically negligible \cite{Ca97}, these
reactions allow to fix `experimentally' the $c\bar{c}$-baryon
dissociation cross section. Furthermore, the Glauber approach for
$c\bar{c}$ absorption in this case should be approximately valid at
energies of about 200 -- 450~GeV such that the transport calculations
can be tested additionally in comparison e.g. to the model calculations
of Refs.~\cite{gerschel,KhaQM96,gavin,gavin1,KhSatz96}.

\begin{figure}
\vspace*{-2cm}
{\psfig{figure=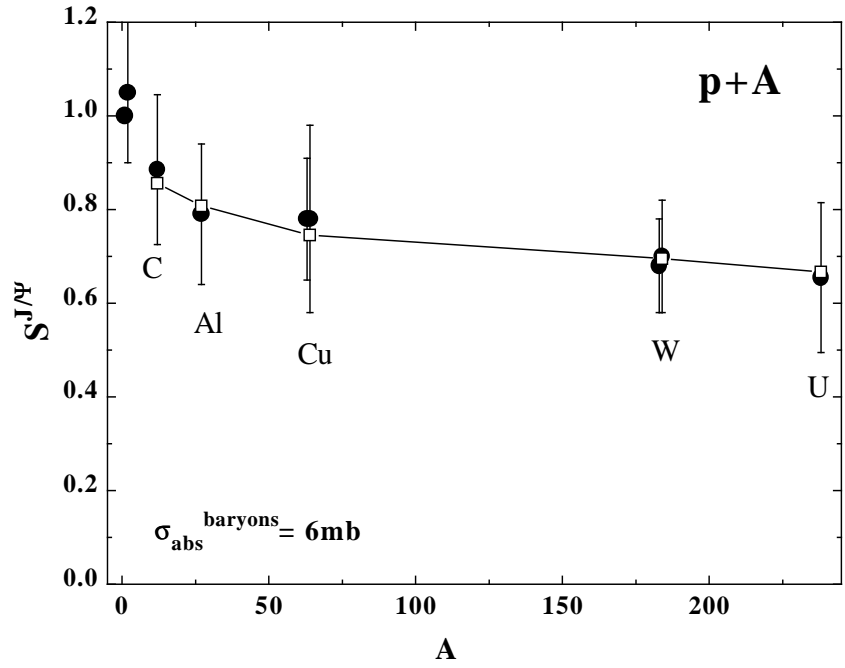,width=15cm,height=22cm}}
\vspace*{-2cm}
\caption{The $J/\Psi$ `survival probability' $S^{J/\Psi}$ for p~+~A
reactions at 200~GeV assuming a 6~mb cross section for the $c\bar{c}$
dissociation on baryons (open squares) in comparison to the experimental
data from \protect\cite{gonin} within the model I.}
\label{Fig8}
\end{figure}

In Fig.~\ref{Fig8} we show the results of our calculations for the
$J/\Psi$ `survival probability' (open squares) using 6~mb for the
absorption cross section of the $c \bar{c}$-pairs on nucleons in
comparison to the data \cite{gonin}. The experimental `survival
probabilities'  $S_{exp}^{J/\Psi}$ in this figure as well as in the
following comparisons are defined by the ratio of experimental
$J/\Psi$ to Drell-Yan cross sections as
\begin{eqnarray}
S_{exp}^{J/\Psi} = \left.{\left(B_{\mu\mu}\sigma^{J/\Psi}_{AB}\over
\sigma^{DY}_{AB}|_{2.9-4.5 \ {\rm GeV}}\right)} \right/
{\left(B_{\mu\mu}\sigma^{J/\Psi}_{pd}\over \sigma^{DY}_{pd}\right)},
\label{sexp}\end{eqnarray}
where $A$ and $B$ denote the target and projectile mass while
$\sigma^{J/\Psi}_{AB}$ and $\sigma^{DY}_{AB}$ stand for the $J/\Psi$ and
Drell-Yan cross sections from $AB$ collisions, respectively, and
$B_{\mu\mu}$ is the branching ratio of $J/\Psi$ to dimuons.
The theoretical ratio is defined as
\begin{eqnarray}
S_{theor}^{J/\Psi} = {M_{J/\Psi}\over N_{J/\Psi}},
\label{stheor}\end{eqnarray}
where $N_{J/\Psi}$ is the multiplicity of initially produced $J/\Psi$'s
while $M_{J/\Psi}$ is the multiplicity of $J/\Psi$'s that survive
the hadronic final state interactions.
We note that due to the large statistical error bars of the
experimental data absorption cross sections $\sigma_{abs}^{baryons}$ of
6 $\pm$ 1~mb are compatible also. These values are slightly smaller
than those of Kharzeev et al.~\cite{KLNS96} in the Glauber model
claiming 7.3~$\pm$~0.6~mb, but in the same range as those used in the
Glauber models of Ref.~\cite{gerschel}. We do not expect to get exactly
the same values as in the Glauber model because the transverse
expansion of the scattered nucleons as well as of the $c\bar{c}$-pairs
is neglected there.  Due to the `optimal' reproduction of the available
data with a cross section of about 6~mb we will use this value also for
nucleus-nucleus collisions in the following.

\subsection{Transverse energy distributions}
\begin{figure}[h1]
\vspace*{-2cm}
{\psfig{figure=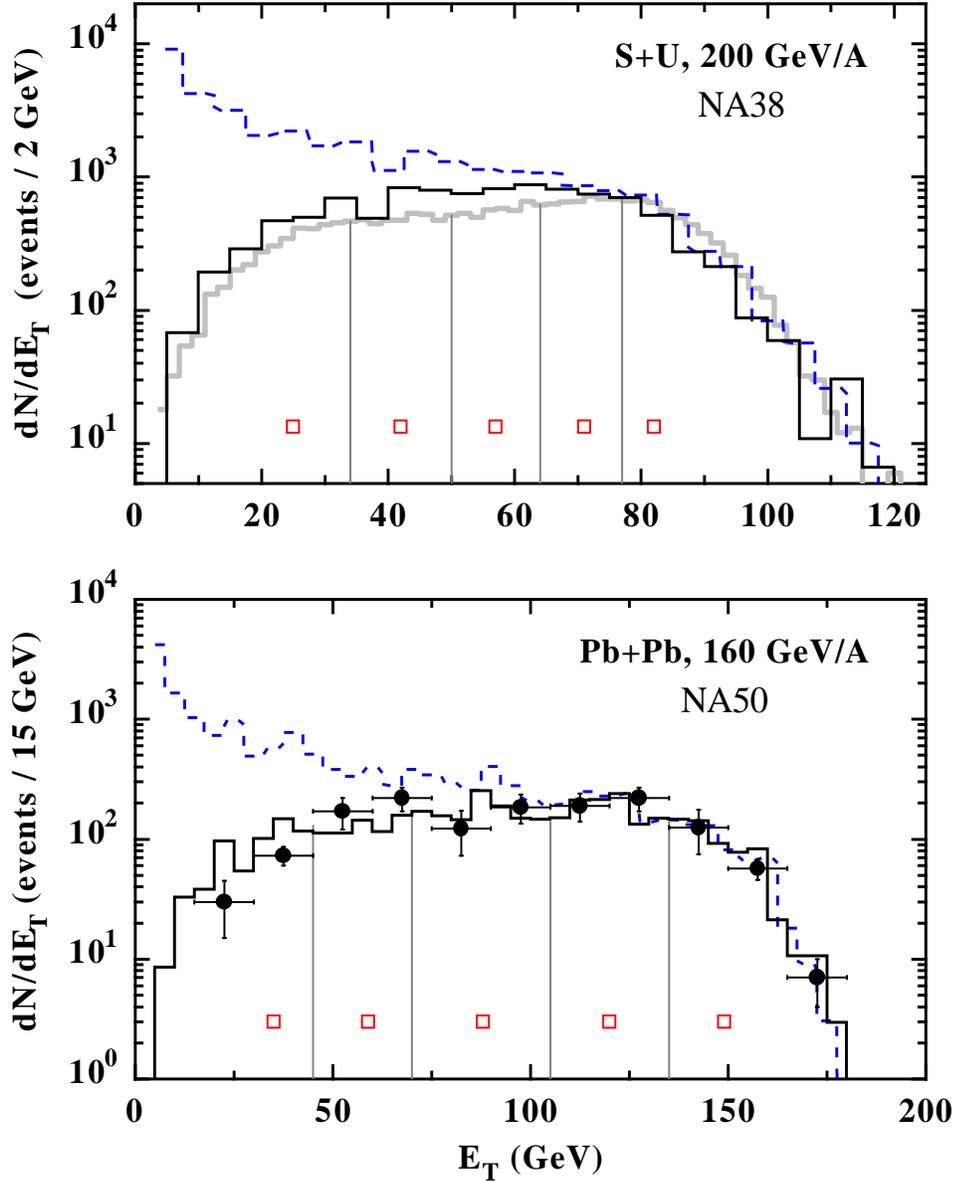,width=15cm,height=22cm}}
\vspace*{-2cm}
\caption{The distributions in the neutral transverse energy for S~+~U
at 200~GeV/A and Pb~+~Pb at 160~GeV/A; dashed histograms: result of the
transport calculations without any constraints on the experimental
acceptance (\protect\ref{dsdet});  experimental distributions: grey
histogram for S + U, full dots for Pb + Pb; solid histograms:
HSD calculation for the $E_T$ distribution according to
Eq.~(\protect\ref{ET-theor}). The open squares denote the average $E_T$
in the experimental $E_T$ bins.}
\label{Fig9}
\end{figure}

In order to perform a detailed comparison to the data of the NA38 and
NA50 Collaborations for S~+~U at 200~GeV/A and Pb~+~Pb at 160~GeV/A we
first have to fix the experimental event classes as a function of the
neutral transverse energy $E_T$ in order to allow for an event by event
analysis. In this respect we compute the differential cross section
\begin{equation}
\frac{d\sigma}{d E_T} = 2 \pi \int_0^\infty b \  db \frac{dN}{d E_T}(b)
\label{dsdet}
\end{equation}
as a function of the impact parameter $b$, where $\frac{d N}{d E_T}(b)$
is the differential $E_T$ distribution for fixed $b$. Since the
detector response is not known to the authors the $E_T$ distribution is
rescaled to reproduce the tail in the experimental $E_T$ distributions
which results in the dashed histograms in Fig.~\ref{Fig9} that increase
for low $E_T$.

In the actual experiments, however, only events are recorded with a
$\mu^+\mu^-$ pair of invariant mass $M\ge 1.5$~GeV and rapidity $3\le
y_{lab} \le 4$ for NA38 or $M\ge 2.9$~GeV and rapidity $2.93\le
y_{lab} \le 3.93$ for NA50. A respective selection in the transport
calculation is obtained by
\begin{eqnarray}
{d\sigma_{theor}^{\mu\mu} \over E_T} = 2\pi N_0 \int\limits_0^\infty
b db \ {dN\over dE_T}(b) \ \sum\limits_{i} W_i^{\mu\mu} (b),
\label{ET-theor}\end{eqnarray}
where $W_i^{\mu\mu} (b)$ are the weights for produced $\mu\mu$ pairs
within the experimental cuts.
$N_0$ is a normalization factor to adjust to the experimental
number of events. The calculated distributions (\ref{ET-theor}) are
shown in Fig.~\ref{Fig9} in terms of the solid histograms which
reasonable reproduce the experimental distributions (grey histogram for
S + U, full dots for Pb + Pb).  The average values for $E_T$ in the 5
bins are indicated in Fig.~\ref{Fig9} by the open squares and coincide
with the corresponding experimental values from Ref.~\cite{NA50}.

\subsection{NA38 and NA50 data}
\begin{figure}[h1]
\vspace*{-2cm}
{\psfig{figure=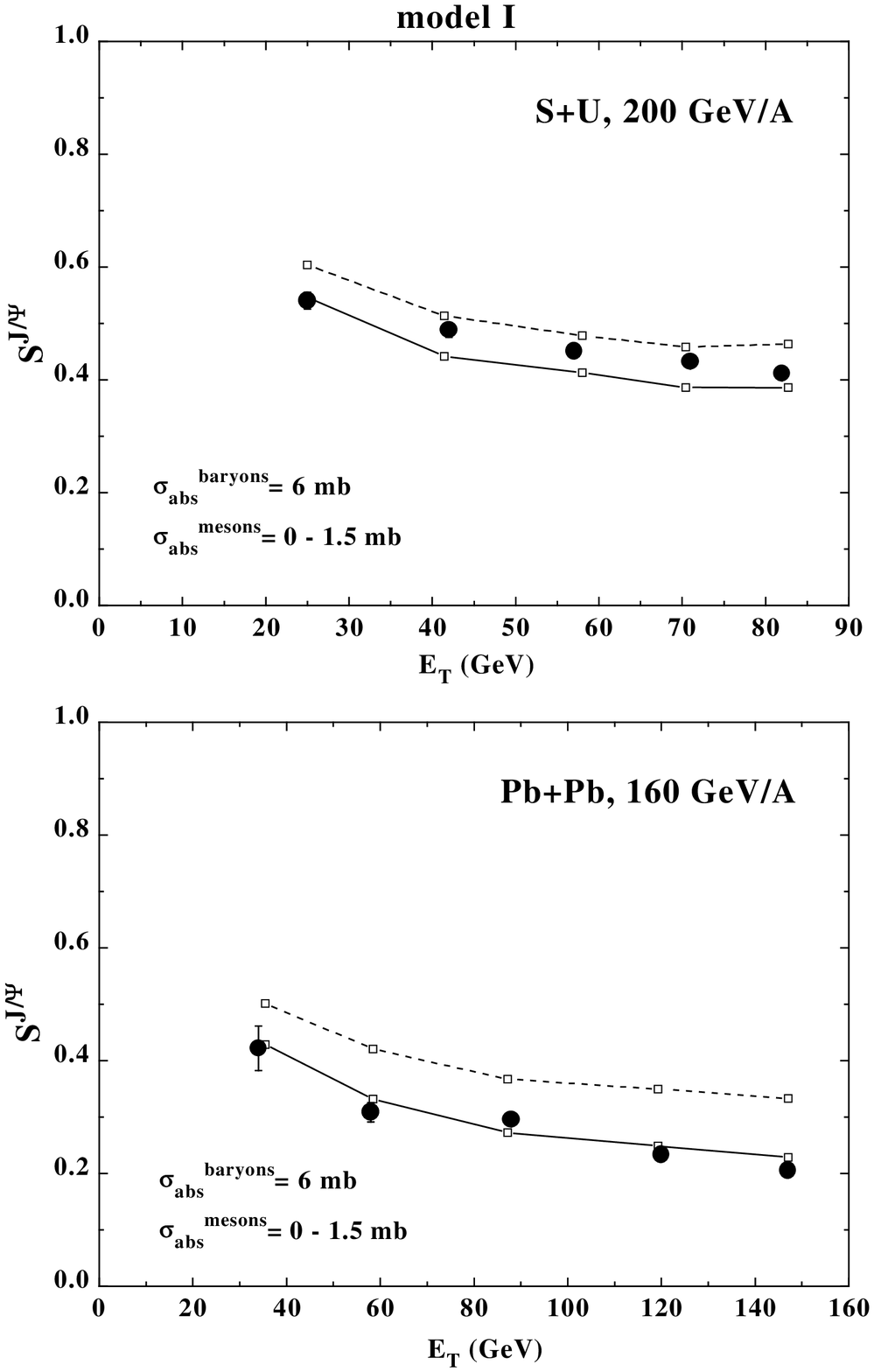,width=15cm,height=22cm}}
\vspace*{-2cm}
\caption{The ratio $S^{J/\Psi}$ for S~+~U at 200~GeV/A (upper
part) and Pb~+~Pb at 160~GeV/A (lower part) as a function of the
transverse energy in comparison to the experimental data from
\protect\cite{NA50} within the model I assuming a long lifetime for
the $c\bar{c}-g$ system.  The absorption cross section on mesons is
varied from 0 (dashed lines) to 1.5~mb (solid lines).}
\label{Fig10}
\end{figure}

We are now in the position to perform a comparison to the experimental
survival probabilities for $J/\Psi$ and $\Psi^\prime$ production in the
respective transverse energy bins. We first compute the results for S +
U at 200~GeV/A and Pb~+~Pb at 160~GeV/A within the model I varying the
dissociation cross section on mesons of the $c\bar{c}-g$ object from 0
to 1.5~mb while keeping the absorption cross section on baryons fixed
at 6~mb.  The calculated $S_{theor}^{J/\Psi}$ are displayed in
Fig.~\ref{Fig10} in comparison to the data for both systems; the dashed
lines are obtained for $\sigma^{mesons}_{abs}$ = 0 mb while the solid
lines correspond to $\sigma^{mesons}_{abs}$ = 1.5 mb.  Whereas the data
for S~+~U appear to be approximately compatible with our calculations
without any dissociation by mesons\footnote{The S+U data are best
described with $\sigma^{baryons}_{abs} \approx$ 6.5 mb.} the Pb~+~Pb
system shows an additional suppression. This finding is in agreement
with the results of Glauber models \cite{KhaQM96,KLNS96}.  On the other
hand, the Pb~+~Pb data are well reproduced with a cross section of
1.5~mb (in model I) for the $J/\Psi$ absorption on mesons which,
however, then slightly overestimates the suppression for the S~+~U data
for the 3 middle $E_T$ bins.  Our calculations thus do not indicate a
strong argument in favor of a QGP phase in the Pb + Pb reaction to
interpret the $J/\Psi$ suppression within the model I.

The latter conclusion is different from the Glauber calculations of
Refs. \cite{KhaQM96,KLNS96} and should be due to the simplified assumptions
about the actual meson abundancy with which the $c\bar{c}$ pair can be
dissociated. Note again that mesons appear in our dynamical approach only
after $\tau_F$ = 0.7 fm/c after the first 'hard' collision.
As has been pointed out by Gavin et al.~\cite{gavin} especially
the $J/\Psi$ in the comoving frame should be dissociated by $\rho$ and
$\omega$ mesons and less by pions due to the large gap in energy for
$D\bar{D}$ dissociation. We have investigated this suggestion more
quantitatively within our microscopic approach and find the following
hadronic decomposition for $J/\Psi$ absorption on mesons: S~+~U
(central):  pions~(35\%), \ $\rho$'s~(42\%), \ $\omega$'s~(15\%),
$\eta$'s~(8\%); \ Pb~+~Pb (central):  pions~(37\%), \ $\rho$'s~(42\%), \
$\omega$'s~(13\%), \ $\eta$'s~(8\%). Thus the hadronic decomposition
practically does not change when going from S~+~U to the heavier Pb~+~Pb
system. As a side remark we note that the dilepton yield from central
S~+~Au and Pb~+~Au collisions -- normalized to the charged particle
multiplicity in the rapidity bin 2.1~$\leq y \leq$~3.1 -- experimentally
appears to be the same \cite{Drees} which is in line with our findings.

\begin{figure}[h1]
\vspace*{-2cm}
{\psfig{figure=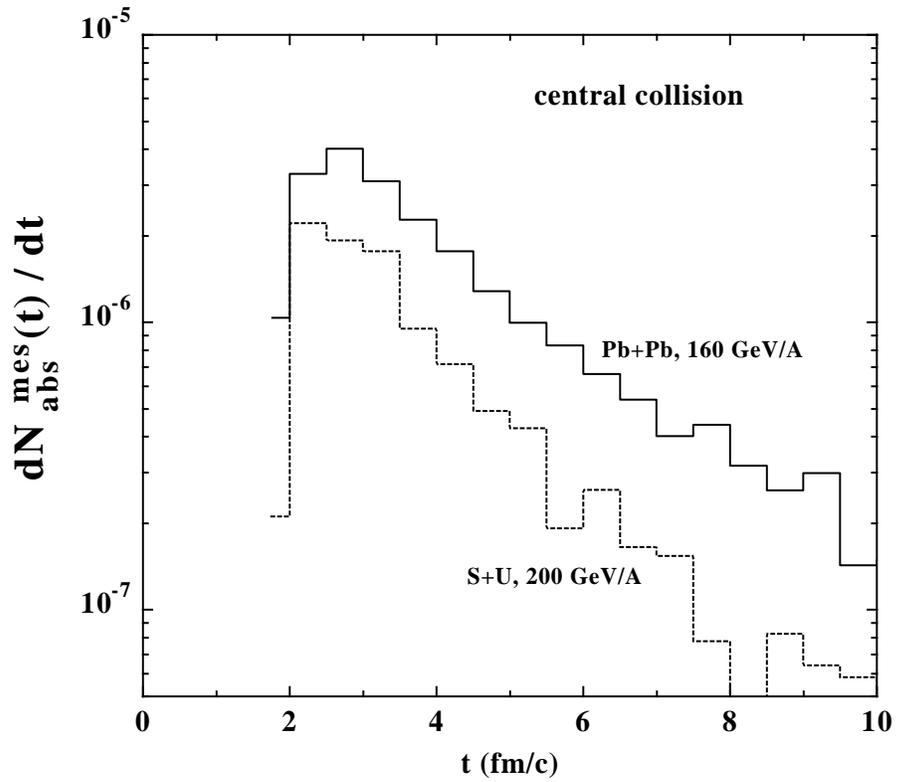,width=15cm,height=22cm}}
\vspace*{-2cm}
\caption{The $J/\Psi$ absorption rate $dN^{mes}_{abs}(t)/dt$ on mesons
for central collisions of S~+~U at 200~GeV/A and Pb~+~Pb at 160~GeV/A. The
absorption rate for S~+~U has been multiplied by a factor $208/32$
to compensate for the different number of projectile nucleons.}
\label{Fig11}
\end{figure}

A remarkable difference, however, is found when comparing the number of
absorbed $J/\Psi$'s by mesons in central collisions for S~+~U and Pb +
Pb as a function of time. In order to compare the two absorption rates
$dN^{mes}_{abs}(t)/dt$ on a relative scale we have multiplied the
absorption rate for the system S~+~U by the ratio of projectile
nucleons ($208/32$) in Fig.~\ref{Fig11}. Here the heavier
system Pb~+~Pb shows higher absorption rates from the beginning (at $t
= 2$~fm/c), which correlates with the central meson densities shown in
Fig.~\ref{Fig2}, and also the $J/\Psi$ absorption on mesons lasts
longer in accordance with the calculations in Ref.~\cite{Ca97}. The
latter effect can also be extracted from the schematic picture in
Fig.~\ref{Fig3} where the number of produced $c\bar{c}$-pairs at finite
meson density is expected to be much larger for Pb~+~Pb than for S~+~U.

\begin{figure}[h1]
\phantom{a}
\vspace*{-2cm}
{\psfig{figure=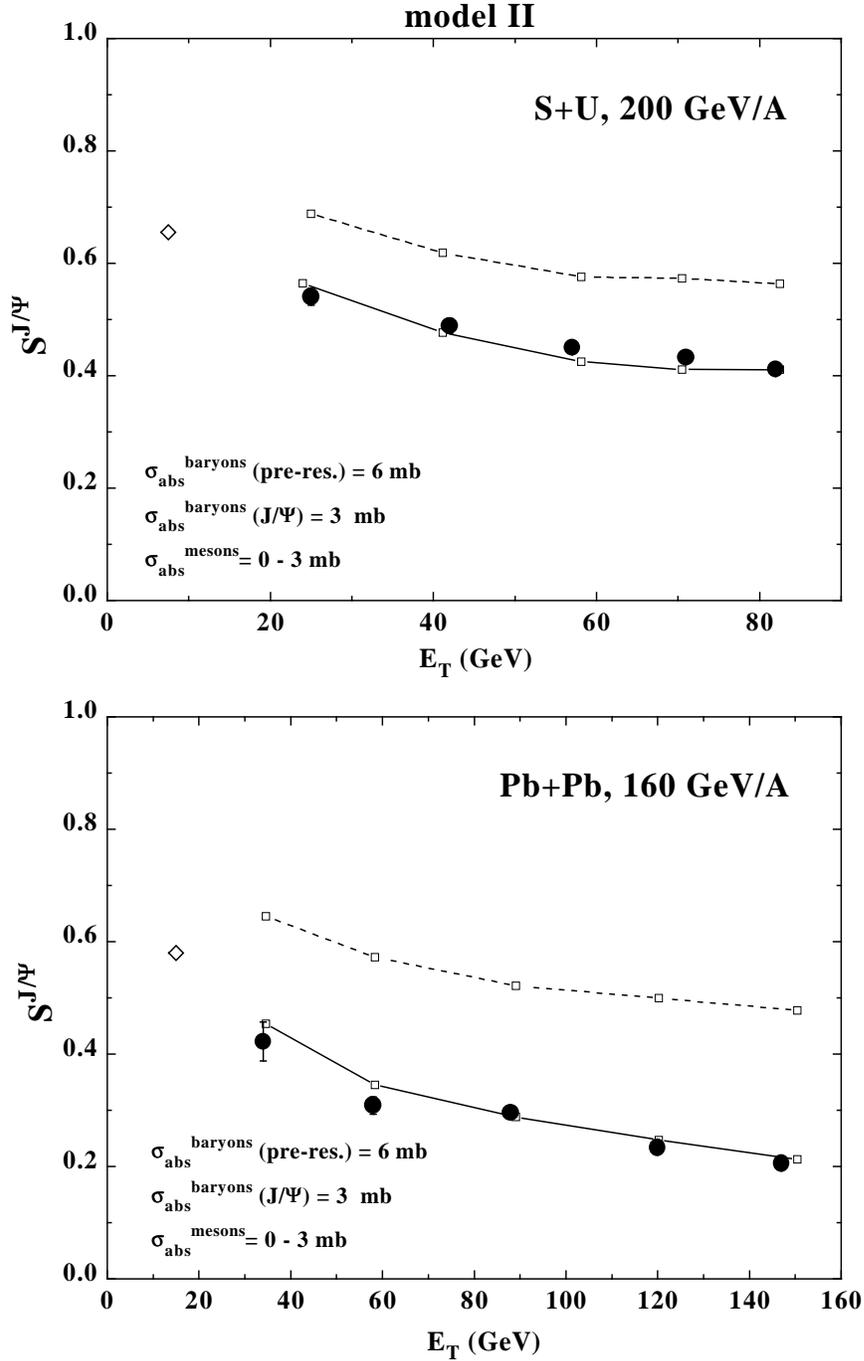,width=15cm,height=22cm}}
\vspace*{-2cm}
\caption{The ratio $S^{J/\Psi}$ for S~+~U at 200~GeV/A
(upper part) and Pb~+~Pb at 160~GeV/A (lower part) as a function of the
transverse energy in comparison to the experimental data from
\protect\cite{NA50} within the model II (see text).  The absorption
cross section on mesons is varied from 0 (dashed lines) to 3~mb (solid
lines); the dissociation cross section on baryons for the pre-resonance
$c\bar{c}-g$ system was taken as 6 mb while for the $J/\Psi$ singlett
cross section with baryons 3 mb were adopted.  The open diamonds
represent the calculated `survival probabilities' $S^{J/\Psi}$
(within model II) for more peripheral reactions (see text).}
\label{Fig12}
\end{figure}

We, furthermore, compute the ratio $S^{J/\Psi}$ for S~+~U at
200~GeV/A and Pb~+~Pb at 160~GeV/A within the model II for a
$c\bar{c}-g$ lifetime $\tau$ = 0.3~fm/c varying the dissociation cross
section of the $c\bar{c}-$pair with mesons from 0 - 3~mb while keeping
the absorption cross section on baryons fixed at $\sigma_{abs}^{baryons}$ =
6~mb for the `pre-resonance' state and at 3~mb for the formed $J/\Psi$
resonance.  The calculated ratios are displayed in
Fig.~\ref{Fig12} in comparison to the data; again the dashed lines
correspond to the calculations without any charmonium absorption on
mesons whereas the solid lines represent our calculations for a meson
absorption cross section of 3~mb. In the absorption model II the data
for S~+~U are no longer compatible with our calculations without any
dissociation by mesons.  The S~+~U data here need an absorption by
mesons in the range of 3~mb as in the phenomenological model of Gavin
et al. \cite{gavin,gavin1}. With $\sigma_{abs}^{mesons} \approx$ 3~mb for
the absorption on mesons, however, the Pb~+~Pb data appear to be
compatible, too. For future experimental studies we display in Fig.
\ref{Fig12} the calculated ratio $S_{theor}^{J/\psi}$ (within model II) for
more peripheral reactions (open diamonds) that correspond to an
$E_T$-bin from $5\div 10$~GeV in case of S + U at 200 GeV/A and to an
$E_T$-bin from $10\div 20$~GeV in case of Pb + Pb at 160 GeV/A.

\begin{figure}[h1]
\vspace*{-2cm}
{\psfig{figure=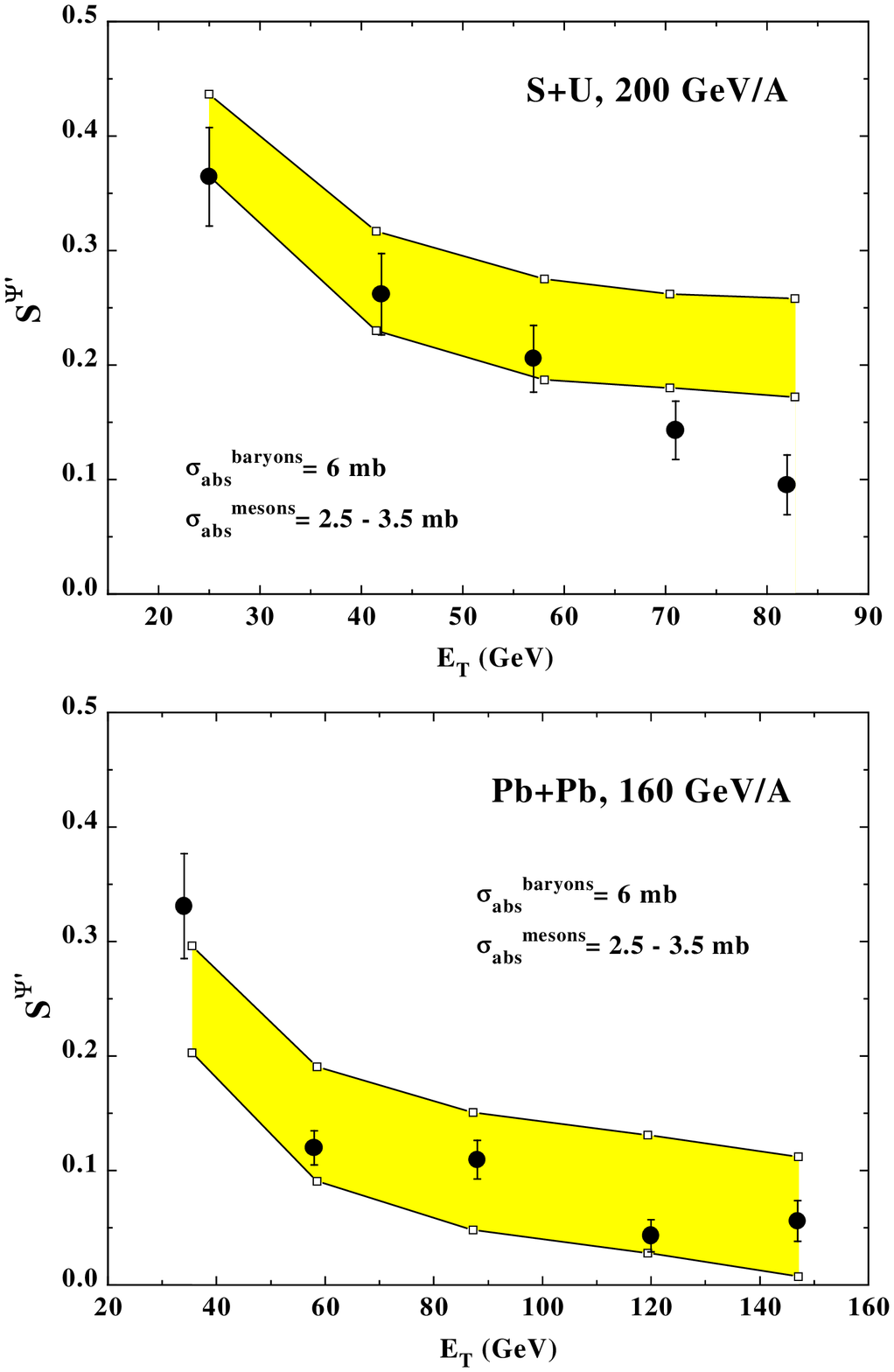,width=15cm,height=22cm}}
\vspace*{-2cm}
\caption{The ratio $S^{\Psi^\prime}$ for S~+~U at 200~GeV/A
(upper part) and Pb~+~Pb at 160~GeV/A (lower part) as a function of the
transverse energy in comparison to the experimental data from
\protect\cite{NA50} within the model I assuming a long lifetime for
the $c\bar{c}-g$ system.  The absorption cross section on mesons is
varied from 2.5 to 3.5~mb.}
\label{Fig13}
\end{figure}

We additionally investigate if the $\Psi^\prime$ suppression measured
by NA38 and NA50 for S~+~U and Pb~+~Pb can be described simultaneously
within our approach without including any additional assumptions.
Since the $\Psi^\prime$ suppression in proton-nucleus reactions is
practically the same \cite{CarQM96} we employ also a dissociation cross
section of 6~mb on baryons while the absorption cross section on mesons
is treated again as a free parameter. Our numerical results for both
systems on the $\Psi^\prime$ survival probability within the absorption
model I are displayed in Fig.~\ref{Fig13} in comparison with the data from 
\cite{gonin}. Due to the higher $\Psi^\prime$ suppression in these
reactions the absorption on mesons requires cross sections from 2.5 --
3.5~mb.  We note that we cannot describe the experimental ratios
$S_{exp}^{\Psi^\prime}$  for central S + U collisions (high $E_T$) within
the present treatment which might be due to the neglect of $\Psi^\prime
\rightarrow J/\Psi + X,$ or $\pi \Psi^\prime \rightarrow J/\Psi + X$
reaction channels.  This also holds for calculations within the model
II where $\sigma_{abs}^{mesons} \simeq 4-5$~mb has to be assumed.
Furthermore, the actual statistics reached for $\Psi^\prime$ production
and absorption events is too low to allow for final conclusions here.

\section{Summary}
In this paper we have carried out a microscopic transport study of
$J/\Psi$, $\Psi^\prime$ and Drell-Yan production in proton-nucleus and
nucleus-nucleus collisions.  Our calculations show that the absorption
of `pre-resonance' $c\bar{c}-g$ states  by both nucleons and produced
mesons can explain reasonably not only the inclusive $J/\Psi$ cross
sections but also the transverse energy ($E_T$) dependence of $J/\Psi$
suppression measured in  nucleus-nucleus collisions.  In particular,
the absorption of $J/\Psi$'s by produced mesons is found to be
important especially for Pb~+~Pb reactions,
where the $J/\Psi$-hadron reactions extend to longer times as
compared to the S~+~W or S~+~U reactions.  This is in contrast with results
based on a simple Glauber model, which neglects both the transverse
expansion of the hadronic system and the finite meson formation times,
where the $c\bar{c}-N$ absorption is roughly sufficient even for
S-induced collisions.  As a consequence we do not find a necessary
argument to require the formation of a quark-gluon-plasma in Pb~+~Pb
collisions at 160 GeV/A. This could only be done through experimental or
theoretical evidence that the charmonium-meson cross sections employed
($\approx $ 1.5 -- 3~mb) in our analysis are too large.  Since the
$c\bar{c}$ dissociation on mesons is expected to be dominated by flavor
exchange reactions the present cross sections in our opinion, however,
should be reasonable.

We close our study by noting that in the phase of high baryon and meson
density both the meson densities (cf. Fig.~\ref{Fig2}) as well as the
associated energy densities \cite{Ca97} are very high such that a
purely hadronic reaction scheme might be questionable. Further
experimental data with good statistics also for p~+~A collisions with
light nuclei are expected to provide more accurate constraints.

\section*{Acknowledgments}
The authors like to acknowledge valuable discussions with L.~Gerland,
C.~Gerschel, C.~Greiner, J.~H\"ufner, D.~Kharzeev, L.~Kluberg, C.~M.~Ko,
C.~Louren{\c c}o, U.~Mosel, C.~Redlich, H.~Satz, O.~V.~Teryaev and
R.~Vogt.  In particular they like to thank L.~Kluberg for sending the
final cross sections from the NA38 and NA50 collaborations from
Ref.~\cite{NA50} and C.~Louren\c co for providing the PYTHIA based
codes that have been implemented in the HSD approach for the
calculation of the Drell-Yan contributions.

\newpage

\end{document}